\documentclass[pra,superscriptaddress,dvipsnames,twocolumn]{revtex4}

\usepackage{layout}
\usepackage{color}
\usepackage[utf8]{inputenc}
\usepackage{amsmath}
\usepackage{tikz}
\usepackage{xcolor}
\usetikzlibrary{arrows}

\usetikzlibrary[topaths]

\usepackage{changes}
\usepackage{amsfonts}
\usepackage{amsmath}
\usepackage{amsthm}
\usepackage{amscd}
\usepackage{amssymb}
\usepackage{amsxtra}
\usepackage{bm}
\usepackage{bbm}         
\usepackage{exscale}
\usepackage{graphics,color}
\usepackage{latexsym}
\usepackage{enumerate}
\usepackage{dcolumn}      
\usepackage{graphics}
\usepackage{epstopdf}
\usepackage[dvips]{epsfig}
\usepackage{graphicx}
\usepackage{hyperref}
\hypersetup{
    colorlinks,%
    citecolor=black,%
    filecolor=black,%
    linkcolor=black,%
    urlcolor=black
}

\newcommand{\ket}[1]{\ensuremath{|#1\rangle}}

\newcommand{\beq}{\begin{equation}}
\newcommand{\eeq}{\end{equation}}

\newcommand{\bdf}{\begin{defn}}
\newcommand{\edf}{\end{defn}}

\newcommand{\tr}{\ensuremath{\text{Tr}}}

\newtheorem{defn}{Definition}

\newcount\mycount
\begin{document}

\title{\textbf{Covert Quantum Communication}}

\author{Juan Miguel Arrazola}
\affiliation{Centre for Quantum Technologies, National University of Singapore, 3 Science Drive 2, Singapore 117543}
\author{Valerio Scarani}
\affiliation{Centre for Quantum Technologies, National University of Singapore, 3 Science Drive 2, Singapore 117543}
\affiliation{Department of Physics, National University of Singapore, 2 Science Drive 3, Singapore 117542.}
\date{\today}

\begin{abstract}
We extend covert communication to the quantum regime by showing that covert quantum communication is possible over optical channels with noise arising either from the environment or from the sender's lab. In particular, we show that sequences of qubits can be transmitted covertly by using both a single photon and a coherent state encoding. We study the possibility of performing covert quantum key distribution and show that positive key rates and covertness can be achieved simultaneously. Covert communication requires a secret key between sender and receiver, which raises the problem of how this key can be regenerated covertly. We show that covert QKD consumes more key than it can generate and propose instead a hybrid protocol for covert key regeneration that uses pseudorandom number generators (PRNGs) together with covert QKD to regenerate secret keys. The security of the new key is guaranteed by QKD while the security of the covert communication is at least as strong as the security of the PRNG.
\end{abstract}

\maketitle

Alice wants to plan a surprise birthday party for Eve, but this is challenging given Eve's notorious eavesdropping skills. Encrypting the invitations she sends to Bob, Charlie, and the other guests may prevent Eve from knowing the content of the messages, but this is not enough: the fact alone that Alice is communicating with her friends will make Eve suspicious, foiling any hopes of a surprise. What Alice needs is a method of communication that is undetectable by Eve: a method for covert communication.

As with other cryptographic tasks, techniques for covert communication date back to ancient times, where messages were hidden in seemingly innocuous objects such as the scalp of travellers, whose hair would be shaved to reveal a hidden message. Modern techniques include classical and quantum steganography \cite{fridrich2009steganography,shaw2011quantum,sanguinetti2016perfectly} and frequency hopping in spread-spectrum radio transmissions \cite{simon1994spread}. Recently, several schemes have been proposed where covert classical communication is achieved by hiding information in the noise of optical channels \cite{bash2013limits,bash2013quantum,che2013reliable,wang2016optimal}. In particular, in Ref. \cite{bash2015quantum}, Bash et al. showed that it is possible to covertly transmit classical information over lossy bosonic channels with thermal noise, even in the presence of a quantum adversary.

In this work, we extend covert communication to the quantum regime by introducing practical protocols to covertly transmit quantum information, using both single photon and coherent state signals. We show that sequences of qubits can be transmitted covertly in the presence of noise originating from the environment or from the sender's lab, giving analytical security bounds for both cases. In the model where noise originates from the lab, security can be obtained even when Eve is given full control of the channel connecting Alice and Bob. This is an improvement with respect to previous work where Eve could not alter the channel parameters. We then study covert quantum key distribution and show that positive key rates and covertness can be achieved simultaneously.

All methods for covert communication require that the parties share a random secret key. This is an important difficulty, as the participants must ensure that they are not detected when they do so and, once they consume the key, they must find ways of covertly regenerating a new one. This is the key regeneration problem in covert communication, which unfortunately has not been previously addressed in the literature. In this work, we first show that covert QKD protocols using sequences of qubits consume more secret key than they produce. We then propose a hybrid approach to key regeneration in which pseudorandom number generators (PRNGs) and covert QKD can be combined to regenerate secret keys. The security of the key is guaranteed by QKD while the security of covert communication can be shown to be at least as strong as the security of the PRNG. 

\textit{Covert qubits.---} Alice wants to transmit a sequence of qubits to Bob in such a way that Eve cannot detect that they are communicating. We assume that Alice is equally likely to communicate or not, and Eve's goal is to correctly distinguish between these two scenarios. Eve's detection error probability $P_e$ is given by $P_e=\frac{1}{2}(P_{FA}+P_{MD})$, where $P_{FA}$ is the probability of a false alarm and $P_{MD}$ is the probability of a missed detection. Alice and Bob's goal is to prevent Eve from performing better than a random guess, i.e. they want that $P_e\geq \frac{1}{2}-\epsilon$ for sufficiently small $\epsilon>0$. We refer to $\epsilon$ as the detection bias.

We consider the case when Alice encodes a qubit state in a single photon across two optical modes. For definiteness, we assume that they correspond to the polarization degree of freedom of a single time-bin mode, but this specification is not required for our results. We further assume that they have access to $N$ such time-bins, each of which may be used to send a qubit signal. Then our protocol for covert quantum communication is simple: For each of the $N$ time-bins, Alice sends a qubit signal with probability $q\ll 1$, and with probability $1-q$, she does nothing. The only thing that changes compared to a regular protocol is that signals are not sent sequentially, but randomly spread out in time. This is an advantage compared to previously proposed protocols for covert classical communication which are more involved \cite{bash2015quantum}. Alice sends on average $Nq$ qubit signals on time-bins that are pre-agreed according to the secret key. For Eve, who doesn't have the key, Alice sends a signal with probability $q$ for each time-bin.
 
\begin{figure}
\begin{center}
\includegraphics[width=0.7\columnwidth]{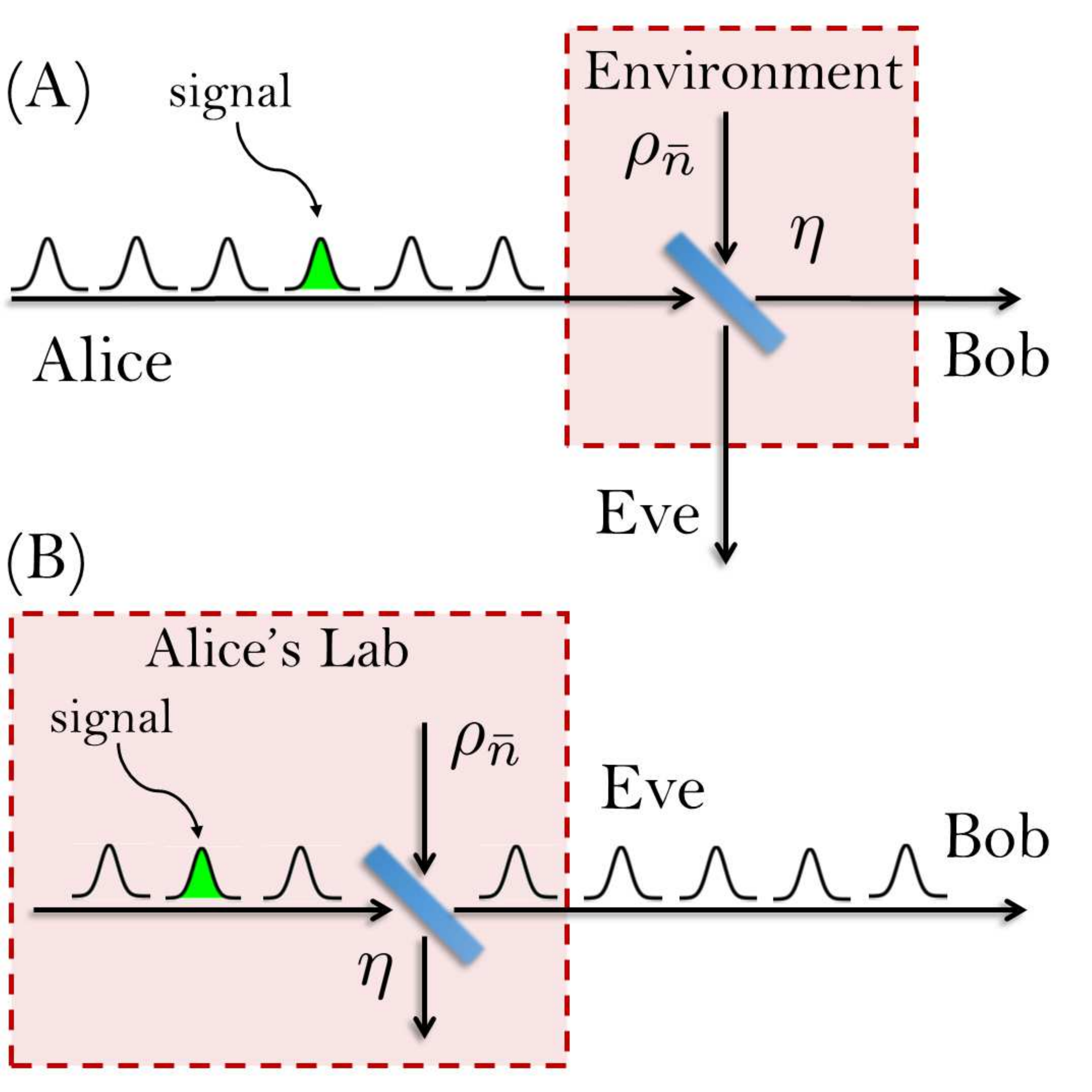}
\caption{Two noise models for covert communication. In case (A), Alice and Bob are connected by a lossy bosonic channel with therml noise, which is modelled as a beam-splitter with transmissivity $\eta$, where the input from the environment is a thermal state $\rho_{\bar{n}}$ with mean photon number $\bar{n}$. In case (B), the therml noise comes from Alice's lab. In both cases, Eve has no control over the parameters inside the red boxes.}\label{Fig:model}
\end{center}
\end{figure} 

For our first noise model, we assume that Alice and Bob are connected by a lossy bosonic channel with thermal noise. This is modelled as a beam splitter with transmissivity $\eta$, where the input from the environment is a thermal state $\rho_{\bar{n}}$ with mean photon number $\bar{n}$, as illustrated in Fig. \ref{Fig:model} (A). As in previous work on covert communication \cite{bash2013limits,bash2013quantum,che2013reliable,bash2015quantum,wang2016optimal}, we assume that Eve has access to all the photons that do not reach Bob, but cannot otherwise change the channel parameters. 

This kind of noise model puts strong constraints on Eve's power, which is not the standard for quantum communication, where Eve is given full control over the channel. To address this issue, we also consider a model where the noise originates from Alice's lab -- which is inaccessible to Eve -- but give Eve full control over the channel connecting Alice and Bob. This model is also illustrated in Fig. \ref{Fig:model} (B).  

For either of the two models, to prove that our protocol is secure, we only need to show that Eve cannot reliably distinguish the $2N$-mode state $\rho$ that she receives when Alice and Bob do not communicate from the state $\sigma$ she receives when they do communicate. The minimum error probability of distinguishing these two states can be bounded as \cite{helstrom76a}
\beq\label{Helstrom}
P_e\geq \frac{1}{2}-\frac{1}{4}||\rho-\sigma||\geq \frac{1}{2}-\sqrt{\frac{1}{8}D(\rho||\sigma)},
\eeq 
where $D(\rho||\sigma)=\tr\left(\rho\log\rho\right)-\tr\left(\rho\log\sigma\right)$ is the relative entropy.

If Alice and Bob do not communicate, the input to the channel is the vacuum state for each polarization mode, and Eve's state is a thermal state. Therefore, Eve's two-mode state $\rho_E$ is 
\beq
\rho_E=\rho_{\bar{n}'}\otimes\rho_{\bar{n}'},
\eeq
where $\bar{n}'=\eta\bar{n}$ in the first model and $\bar{n}'=(1-\eta)\bar{n}$ in the second model. On the other hand, when Alice and Bob communicate, Eve's two-mode state is
\beq
\sigma_E=(1-q)\rho_E+q\rho_s
\eeq
where, as before, $\rho_E$ is Eve's state when there is no signal and $\rho_s$ is her state when Alice sends a qubit signal, which can be calculated for both of our models (see Supplementary Material). Since Alice and Bob independently choose whether to send a signal or not for each time-bin, Eve's $2N$-mode states $\rho$ and $\sigma$ are
tensor product states of the form $\rho=\left(\rho_E\right)^{\otimes N}$, $\sigma=\left(\sigma_E\right)^{\otimes N}$.

From Eq. \eqref{Helstrom}, the detection bias $\epsilon$ can be bounded in terms of the relative entropy between these states as
\begin{align}
\epsilon\leq \sqrt{\frac{1}{8}D(\rho||\sigma)}\nonumber=\sqrt{\frac{N}{8}D(\rho_E||\sigma_E)},
\end{align}
where we have used the fact that the relative entropy is additive for tensor product states. 

In general, Alice may send different qubit states in her signals, each of which would lead to a different state for Eve. In such cases, we can bound the detection bias by considering only the worst case among all signal states
\beq\label{Eq:epsilon}
\epsilon\leq \max_i \sqrt{\frac{N}{8}D(\rho_E||\sigma_{E,i})}
\eeq
where $\sigma_{E,i}$ is Eve's state when Alice sends the $i$-th state. Thus, from now on we simply assume that $\sigma_{E}$ corresponds to the worst-case signal. 

For given values of the parameters $N, \bar{n}, \eta$, and $q$, we can use Eq. \eqref{Eq:epsilon} to bound $\epsilon$ and quantify the security of the protocol. However, we are also interested in analytical bounds that showcase the role of these parameters explicitly. We assume that Alice and Bob want to send an average of $d$ qubit signals, which fixes $N$ and $q$ to satisfy $Nq=d$. For both of our models, we can upper bound the relative entropy through a Taylor series expansion, keeping only terms to second order in $q$. We then introduce an additional bound over values of the transmissivity $\eta$, which notably leads to the same bound for both our models (See Supplemental Material). The resulting bound on the detection bias is
\beq
\epsilon\leq\sqrt{\frac{d\bar{n}^2}{8(1+\bar{n})^3}+\frac{1+4\bar{n}+5\bar{n}^2+3\bar{n}^3}{16\bar{n}(1+\bar{n})^3}\frac{d^2}{N}}.
\eeq
In the regime where $\bar{n}\ll 1$ and $q>\bar{n}^2$, we can approximate this bound as
\beq\label{Eq: scaling}
\epsilon\lesssim \frac{d}{4}\sqrt{\frac{1}{\bar{n}N}},
\eeq
which gives us a $\frac{1}{\sqrt{N}}$ scaling of the detection bias as a function of the number of time-bins $N$ in this regime. From this expression we can also deduce a $\frac{1}{\sqrt{N}}$ scaling for the number of covert qubits $d$ that can be transmitted for fixed $\epsilon$. Furthermore, the mean photon number $\bar{n}$ places a limit to how small the detection bias can be, since the upper bound can never be smaller than $\sqrt{\frac{d\bar{n}^2}{8(1+\bar{n})^3}}\approx\sqrt{\frac{d}{8}}\bar{n}$.

\textit{Covert communication with coherent states.---} Although the polarization of a single photon defines a qubit, in practical implementations, it is usually more convenient to use coherent states. Instead of a single-photon qubit state $\ket{\psi}=\lambda_1\ket{1}_H+\lambda_2\ket{1}_V$, where $\ket{1}_H$ and $\ket{1}_V$ correspond to a single photon in the horizontal and vertical polarization modes respectively, we employ the state 
\beq
\ket{\alpha,\psi}=\ket{\alpha\lambda_1}_H\otimes\ket{\alpha\lambda_2}_V,
\eeq
where $|\alpha|^2=\mu$ is the total mean photon number \cite{arrazola2014quantum}. Notice that in this case we have a product state of both polarization modes.

We require that the average number of photons received by Bob when using coherent states is the same as in the single photon case. This means that if $q$ is the probability of selecting a time-bin in the single photon case and $q'$ is the corresponding probability in the coherent state case, we must have $q=\mu q'$. 

As before, we define $\sigma_{E}$ to be Eve's state when Alice and Bob send a worst-case signal, which can be straightforwardly computed (see Supplemental Material). We can bound the detection bias as in Eq. \eqref{Eq:epsilon}, with a further bound on the relative entropy by using a Taylor series expansion to second order in $q$ and $\mu$, while also bounding over values of $\eta$ (see Supplemental Material). This leads to the expression
\begin{align*}
\epsilon\leq& \left[\left(\frac{\bar{n}^2(1+2\bar{n})}{\mu(1+\bar{n})^4}+\frac{\bar{n}}{(1+\bar{n})^2}+\frac{\bar{n}\mu}{3(1+\bar{n})^2}\right)\right.\frac{d}{8}\\
+&\left.\left(\frac{\bar{n}(1+2\bar{n})+2\bar{n}}{2\mu^2(1+\bar{n})^4}+\frac{1+7\bar{n}+16\bar{n}^2+12\bar{n}^3}{2\bar{n}(1+\bar{n})^3}\right)\frac{d^2}{8N}\right]^{\frac{1}{2}}.
\end{align*}
For $\bar{n}\ll 1$ and $\mu\ll 1$ with $\mu\gg\bar{n}$, we can approximate this bound as
\beq\label{scaling coherent}
\epsilon\lesssim \sqrt{\frac{d^2}{16N\bar{n}}+\frac{d\bar{n}}{8}},
\eeq
which also gives us a $\frac{1}{\sqrt{N}}$ scaling for the detection bias. 
\begin{figure}
\begin{center} 
\includegraphics[width=0.8\columnwidth]{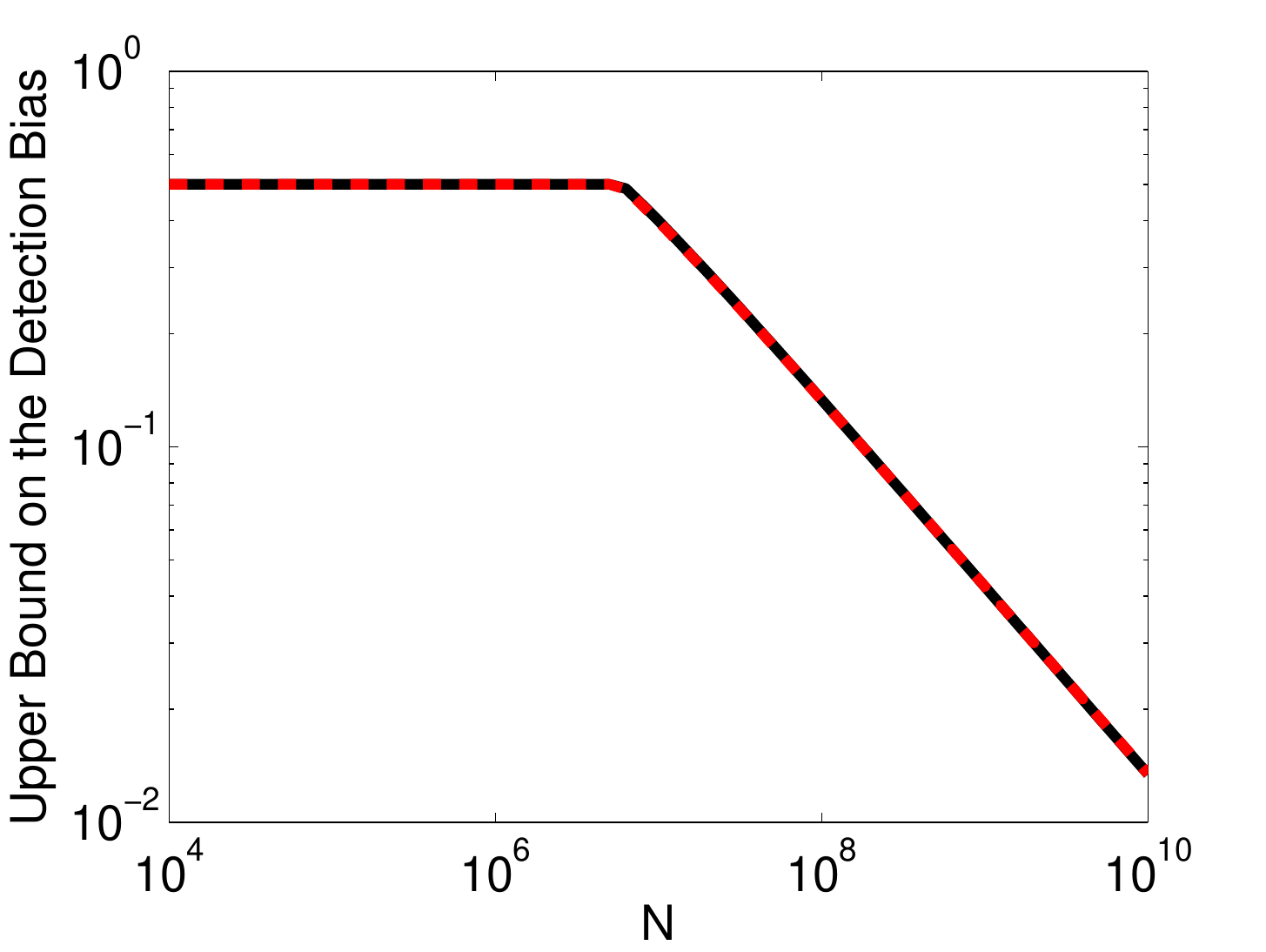}
\caption{Log-log plot of the upper bound of Eq. \eqref{Eq:epsilon} as a function of the number of time-bins $N$.  We consider the model where noise comes from Alice's lab. The black curve corresponds to the single photon protocol and the dashed red curve to the coherent state protocol. Both curves overlap almost perfectly. We have set $d=20$, $\eta=0.5$, and $\bar{n}=10^{-5}$, as would be the case for a thermal state at temperature $300$K and infrared wavelength of 4.2$\mu$m. The mean photon number used in the coherent state case is $\mu=10^{-3}$.}\label{Fig:Plot}
\end{center}
\end{figure} 

Figure \ref{Fig:Plot} illustrates the behaviour of the detection bias when Alice and Bob use single-photon and coherent-state signals in the model where noise comes from Alice's lab. The square-root scaling of the detection bias can be clearly seen in both cases for a certain range of values of $N$. Overall, both implementations behave similarly and for the values considered, can achieve a detection bias below 2\%. 

\textit{Covert QKD.---} Based on these results, we study the possibility of performing covert quantum key distribution (QKD). Two issues must be addressed. First, the possibility of keeping the classical post-processing covert, which can be carried out using existing protocols for covert classical communication \cite{bash2015quantum}. Most QKD protocols require two-way classical communication, which can be achieved in our first model since the situation is symmetric for both Alice and Bob, while for the second model, it requires Bob to also have a source of noise in his lab.

Second, we need to prove that Bob can actually use the weak signals sent by Alice. To do this, we show that, on the channel with parameters that guarantee low detection bias $\epsilon$, Alice and Bob would have positive key rate even if all the errors were attributed to Eve. For definiteness, we focus on the BB84 protocol with both a single photon and coherent state implementation. The asymptotic key rates, with optimal error-correction, are respectively \cite{scarani2009security}
\begin{align}
K_S&=R(1-2h(Q))\label{KeyRateSP}\\
K_C&=R[Y_1(1-h(Q/Y_1))-h(Q)]\label{KeyRateCoherent}
\end{align}
where $h(\cdot)$ is the binary entropy, $R$ is the total detection rate, $Q$ is the quantum bit error rate, $Y_1=\max(0,1-\frac{\mu}{2\tau})$, and $\tau$ is the total transmissivity. In general, Eve has control over the error rate so she can always prevent Alice and Bob from establishing a key. What we need to show is that they can achieve positive key rates despite the presence of the noise that is required for covert communication. This will occur if the resulting error rates $Q$, arising solely from the noise, are sufficiently low.

In the model where noise comes from the lab and in the absence of additional experimental imperfections, the error rates due to noise in the single photon and coherent state case respectively satisfy (see Supplemental Material) 
\begin{align}
Q_S\approx \left(\frac{1}{\eta}-1\right)\bar{n},\hspace{0.5cm}Q_C\approx\left(\frac{1}{\eta}-1\right)\frac{\bar{n}}{\mu}.\label{QBER} 
\end{align}
In the single photon case, the key rate of Eq. \eqref{KeyRateSP} is positive as long as $Q_S<0.11$, which can be easily achieved whenever $\bar{n}\ll 1$.  Similarly, in the coherent state case, positive key rates can be obtained if $\mu\leq\tau$ and $Q_C\ll 1$, which occurs whenever $\bar{n}\ll \mu$. Thus, we can have positive key rates for covert QKD even in the presence of noise. In order to also achieve a small detection bias, we must simply set a sufficiently large number of time-bins $N$. From Eqs. \eqref{Eq: scaling} and \eqref{scaling coherent}, this requires setting $N\gg\frac{d^2}{\bar{n}}$ in both the single photon and the coherent state case.
 
For the values used in Fig. \ref{Fig:Plot}, which are $\bar{n}=10^{-5}$, $\mu=10^{-3}$, and $\eta=0.5$, we can obtain key rates of $K_S= 0.99 R$, and $K_C= 0.47 R$ for transmissivity $t=\mu$, while still achieving a detection bias smaller than $2\%$ for $N\sim 10^{10}$ time-bins. This shows that it is possible to simultaneously achieve non-zero key rates and a low detection bias in covert QKD. For other quantum communication protocols, it should also suffice to set $\bar{n}\ll 1$ in the single photon case and $\bar{n}\ll \mu$ in the coherent state case, since these conditions imply a large signal to noise ratio for the receiver. 

As we argue in the next section, an application of covert QKD is that it allows Alice and Bob to regenerate the secret strings that are required for covert communication.

\textit{Secret key regeneration.---} Ideally, we would like to run a covert QKD protocol that generates more secret key than it consumes. However, this is not possible to achieve with QKD protocols that use a sequence of qubits as signals. In a covert QKD protocol, Alice independently sends a signal with probability $q$ for each of the $N$ available time-bins. In the limit of large $N$, the average amount of shared bits needed to specify the selected time-bins is $N\,h(q)$, where $h(\cdot)$ is the binary entropy. On the other hand, at best, Alice and Bob only obtain an average of $d=Nq$ secret bits from running the covert QKD protocol, without including the overhead required by parameter estimation and error-correction. Since $h(q)>q$ for $q<\frac{1}{2}$, it follows necessarily that such a protocol for covert QKD consumes more key than it can produce. We are thus forced to look for alternatives. We propose instead a hybrid key regeneration method, where Alice and Bob obtain computational security for the covert communication by using pseudorandom number generators (PRNGs) to decide for which time-bins they send their signals, while retaining information-theoretic security of the regenerated secret key. 

PRNGs take a truly random key as input and expand it into an exponentially larger pseudorandom output. The PRNG is secure if this output cannot be distinguished from a truly random string. Since the pseudorandom output is much larger than the seed, it can now be used to perform large amounts of covert QKD, thus generating a new secret key larger than the original key that was used as input, allowing an indefinite amount of key regeneration \footnote{For example, for $q=10^{-10}$, the ratio $h(q)/q$ (which increases for small $q$) is approximately 35, while using standard PRNGs like CTR-DRBG (AES), one can obtain at least $2^{19}\approx 5\times 10^5$ pseudorandom bits per call from a 440-bit input string according to security recommendations appearing in NIST SP 800-90A.}. The security of the QKD protocol is unchanged during covert communication, so the new key will have information-theoretic security. On the other hand, the security of the covert communication can be shown to be at least as strong as the security of the underlying PRNG.

To see this, assume that Eve can break the security of the covert communication protocol implemented with a PRNG output, but she can't if it is implemented with a truly random string. Then Eve has the power to break the security of the PRNG too. Indeed, provided with the alleged random string, she could simply run the covert communication protocol herself: if she can't break its security, the string must be truly random and if she can, the string must be pseudorandom. By contrapositive, this implies that if Eve cannot break the security of the PRNG, in particular she cannot break the security of a covert communication protocol that uses a PRNG output instead of true randomness. Thus, the security of covert communication is at least as strong as the security of the PRNG.

\textit{Discussion.---} We have given simple methods to perform covert quantum communication over noisy optical channels. In the case of QKD, we have shown that small detection biases and positive key rates can be obtained simultaneously. For implementations, it will be crucial to choose optimal wavelengths that lead to significant noise that is still low enough for Bob to reliably detect the signals. We note that in principle any type of optical noise can be used, not just thermal noise. It will also be important to determine the conditions under which the large running times required for covert QKD will be manageable in a practical setting. Although in this work we have focused on QKD, it will be interesting to apply our techniques for covert quantum communication to other protocols. 

\textit{Acknowledgements.---} J.M. Arrazola would like to thank A. Ignjatovic for valuable discussions. This work is funded by the Singapore Ministry of Education (partly through the Academic Research Fund Tier 3 MOE2012-T3-1-009) and the National Research Foundation of Singapore, Prime Minister’s Office, under the Research Centres of Excellence programme.

\bibliography{Bibliography}
\bibliographystyle{apsrev}

\end{document}